%% file: main.tex
\documentclass[sigconf,9pt]{acmart}
\usepackage[utf8]{inputenc}
\usepackage{hyperref}
\hypersetup{pdfstartview=FitH,pdfpagelayout=SinglePage}
\usepackage{xspace}
\usepackage{xcolor}
\usepackage{subcaption}
\usepackage{comment}
\usepackage{graphicx}

\usepackage{tikz}
\usepackage{amsmath}
\usepackage{caption}
\usepackage{multirow}
\usepackage{boldline}
\usepackage[inline]{enumitem}
\usepackage[ruled,vlined]{algorithm2e}
\usepackage{tabularx}
\usepackage{pythonhighlight}

\renewcommand\footnotetextcopyrightpermission[1]{} 
\begin{document}
\title{WiFi Physical Layer Stays Awake and Responds \\When it Should Not}

\author{Ali Abedi}
\affiliation{
\institution{Stanford University}
\country{USA}}
\email{abedi@stanford.edu}

\author{Haofan Lu}
\affiliation{
\institution{UCLA}
\country{USA}}
\email{haofan@cs.ucla.edu}

\author{Alex Chen}
\affiliation{
\institution{University of Waterloo}
\country{Canada}}
\email{zihanchen.ca@gmail.com}

\author{Charlie Liu}
\affiliation{
\institution{University of Waterloo}
\country{Canada}}
\email{charlie.liu@uwaterloo.ca}

\author{Omid Abari}
\affiliation{
\institution{UCLA}
\country{USA}}
\email{omid@cs.ucla.edu}

\begin{abstract}
WiFi communication should be possible only between devices inside the same network. However, we find that all existing WiFi devices send back acknowledgments (ACK) to even fake packets received from unauthorized WiFi devices outside of their network. Moreover, we find that an unauthorized device can manipulate the power-saving mechanism of WiFi radios and keep them continuously awake by sending specific fake beacon frames to them. Our evaluation of over 5,000 devices from 186 vendors confirms that these are widespread issues. We believe these loopholes cannot be prevented, and hence they create privacy and security concerns.  Finally, to show the importance of these issues and their consequences, we implement and demonstrate two attacks where an adversary performs battery drain and WiFi sensing attacks just using a tiny WiFi module which costs less than ten dollars.  
\end{abstract}
\maketitle
\pagestyle{plain} 

\input{introv2.tex}

\input{Related.tex}

\input{polite-wifi}

\input{keep-awake}

\input{Implicaiton_Sense.tex}

\input{Implicaiton_Battery.tex}

\input{Conclusion}


\bibliographystyle{ACM-Reference-Format}
\bibliography{main}
\end{document}

%% file: introv2.tex
\section{Introduciton}

Today's WiFi networks use advanced authentication and encryption mechanisms (such as WPA3) to protect our privacy and security by stopping unauthorized devices from accessing our devices and data. Despite all these mechanisms, WiFi networks remain vulnerable to attacks mainly due to their physical layer behaviors and requirements defined by WiFi standards. In this paper, we find two loopholes in the IEEE 802.11 standard for the first time and show how they can put our privacy and security at risk. 

\textbf{a) WiFi radios respond when they should not.}  In a WiFi network, when a device sends a packet to another device, the receiving device sends an acknowledgment back to the transmitter. In particular, upon receiving a frame, the receiver calculates the cyclic redundancy check (CRC) of the packet in the physical layer to detect possible errors. If it passes CRC, then the receiver sends an Acknowledgment (ACK) to the transmitter to notify the correct reception of the frame. Surprisingly, we have found that all existing WiFi devices send back ACKs to even fake packets received from unauthorized WiFi devices outside of their network. Why should a WiFi device respond to a fake packet from an unauthorized device?! 

\textbf{b) WiFi radios stay awake when they should not.}
WiFi chipsets are mostly in sleep mode to save power. However, to make sure that they do not miss their incoming packets, they notify their WiFi access point before entering sleep mode so that the access point buffers any incoming packets for them. Then, WiFi devices wake up periodically to receive beacon frames sent by the associated access point. In regular operation, only the access point sends beacon frames to notify the devices that have buffered packets. When a device is notified, it stays awake to receive them. However, these beacon frames are not encrypted. Hence, we find that an unauthorized user can forge those beacon frames to keep a specific device awake for receiving the (non-existent) buffered frames. 

We examine these behaviors and loopholes in detail over different WiFi chipsets from different vendors. Our examination of over 5,000 WiFi devices from 186 vendors shows that these are widespread issues. We then study the root cause of these issues and show that, unfortunately, they cannot be fixed by a simple solution such as updating WiFi chipsets firmware.  Finally, we implement and demonstrate two attacks based on these loopholes. In the first attack, we show that by forcing WiFi devices to stay awake and continuously transmit, an adversary can continuously analyze the signal and extract personal information such as the breathing rate of the WiFi users. In the second attack, we show that by forcing WiFi devices to stay awake and continuously transmit, the adversary can quickly drain the battery, and hence disable WiFi devices such as home and office security sensors. These attacks can be performed from outside buildings despite the WiFi network and devices being completely secured. All the attacker needs is a \$10 microcontroller with integrated WiFi (such as ESP32) and a battery bank. The attacker device can easily be carried in a pocket or hidden somewhere near the target building. 

The main contributions of this work are:
\begin{itemize}
    \item We find that WiFi devices respond to fake 802.11 frames with ACK, even when they are from unauthorized devices. We also find that WiFi radios can be kept awake by sending them fake beacon frames indicating they have packets waiting for them. 
    \item We study these loopholes and their root causes in detail, and have tested more than 5,000 WiFi access points and client devices from more than 186 vendors.  
    
    \item We implement two attacks based on these loopholes using just a 10-dollar off-the-shelf WiFi module and validate them in real-world settings.

\end{itemize}

%% file: Related.tex
\section{Related Work}
The loopholes we present in this paper are explored using packet injection, in which an attacker sends fake WiFi packets to devices in a secured WiFi network.
Packet injection has been used in the past to perform various types of attacks against WiFi networks
such as denial of service attacks for a particular client device or total disruption of the network~\cite{vanhoef2020protecting, dos, rogue-ap, deauth}. These attacks use different approaches such as beacon stuffing to send false information to WiFi devices~\cite{beacon-stuffing-1, beacon-stuffing-2}, or Traffic Indication Map (TIM) forgery to prevent clients from receiving data ~\cite{bellardo2003802, tim-forgery}. However, all of these attacks focus on spoofing 802.11 MAC-layer management frames to interrupt the normal operation of WiFi networks. 
To provide a countermeasure for some of these attacks, the 802.11w standard~\cite{ieee802.11w} 
introduces a protected management frame that prevents attackers from spoofing 802.11 management frames. 
Instead of spoofing 802.11 MAC frames, we exploit properties of the 802.11 physical layer to force a device to stay awake and respond when it should not. 
These loopholes open the door to multiple research avenues including new security and privacy threats.

\textbf{WiFi sensing attack:} Over the past decade, there has been a significant amount of research on WiFi sensing where WiFi signals are used to detect human activities~\cite{iot-wifi-localization, rf-sensing, wifi-sensing-survey,adib2015smart, breathing-rate-1, breathing-rate-2,gesture-recognition-1, gesture-recognition-2, gesture-recognition-3,pu2013whole}. However, these systems target applications with social benefits and cannot be easily used by an attacker to create privacy and security threats. This is because either these techniques require cooperation from the target WiFi device or the attacker needs to be very close to the target to use these systems. A recent study shows that by capturing WiFi signals coming out of a private building, it is possible for an adversary to track user movements inside that building~\cite{zhu2018tu}. However, this attack has a bootstrapping stage which requires the attacker to walk around the target building for a long time to find the location of the WiFi devices. Furthermore, since this work relies on only the normal intermittent WiFi activities, it cannot capture continuous data such as breathing rate.  

\textbf{Battery draining attack:} 
Battery draining attacks date back to 1999 \cite{stajano1999resurrecting} and there have been many studies on such attacks and potential defense mechanisms since then~\cite{caviglione2012energy}.
Battery discharge models and energy vulnerability due to operating systems have been investigated \cite{zhang2010accurate,jindal2013hypnos}. A more recent study plays multimedia files implicitly to increase power consumption during web browsing \cite{fiore2014multimedia, fiore2017exploiting}. In terms of defending, a monitoring agent that searches for abnormal current draw is discussed in \cite{buennemeyer2008mobile}. In contrast, our attack exploits the loopholes in the 802.11 physical layer protocol and the power-hungry WiFi transmission to quickly drain a target device's battery. We will discuss in Section~\ref{sec:cannot-be-fixed} that stopping our proposed attack is nearly impossible on today's WiFi devices.

This paper is an extension of our previous workshop publication ~\cite{polite-wifi}. The workshop paper shows preliminary results for our finding that WiFi devices respond with ACKs to packets received from outside of their network, and provides a brief discussion on potential privacy and security concerns of this behavior without studying them. We have also explored how the WiFi power saving mechanism can be exploited to keep a target device awake in a localization attack~\cite{wi-peep}. 
In this paper, we provide an in-depth study of these previously discovered loopholes. We also design and perform two privacy and security attacks, based on these loopholes. Finally, we implement these attacks on off-the-shelve WiFi devices and present detailed performance evaluations.

%% file: polite-wifi.tex
\section{WiFi Responds When It Should Not}\label{sec:polite-wifi}

Most networks use security protocols to prevent unauthorized devices from communicating with their devices. Therefore, one may assume that a WiFi device only acknowledges frames received from the associated access point or other devices in the same network. However, we have found that all today's WiFi devices acknowledge even the frames they receive from an unauthorized device from outside of their network. In particular, as long as the destination address matches their MAC address, their physical layer acknowledges it, even if the frame has no valid payload. In this section, we examine this behavior in more detail, and explain why this problem happens and why it is not preventable.

\begin{figure}[!t]
    \centering
    \includegraphics[width = 0.8\columnwidth]{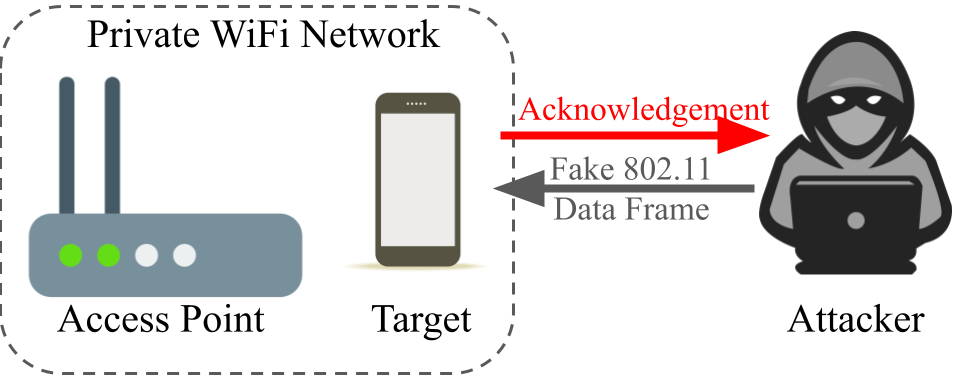}
    \caption{WiFi devices send an ACK for any frame they receive without checking if the frame is valid.}
    \label{fig:polite-wifi}
\end{figure}

To better understand this behavior, we run an experiment where we use two WiFi devices to act as a victim and an attacker. The attacker sends fake WiFi packets to the victim. We monitor the real traffic between the attacker and the victim's device.

\vspace{0.05in}
\noindent \textbf{Setup:} For the victim, we use a tablet, and for the attacker, we use a USB WiFi dongle that has a Realtek RTL8812AU 802.11ac chipset. This is a \$12 commodity WiFi device. The attacker uses this device to send fake frames to the victim's device. To do so, we develop a python program that uses the Scapy library~\cite{scapy} to create fake frames. Scapy is a python-based framework that can generate arbitrary frames with custom data in the header fields. Note, that the only valid information in the frame is the destination MAC address (i.e., the victim's MAC address). The transmitter MAC address is set to a fake MAC address (i.e., aa:bb:bb:bb:bb:bb), and the frame has no payload (i.e., null frame) and is not encrypted.

\begin{figure}[t!]
        \centering
        \includegraphics[width=0.8\linewidth, page=2]{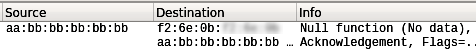}
        \caption{Frames exchanged between attacker and victim} 
        \label{fig:wireshark}
\end{figure}

\vspace{0.05in}
\noindent
\textbf{Result:} Figure~\ref{fig:wireshark} shows the real traffic between the attacker and the victim device captured using Wireshark packet sniffer~\cite{wireshark}. As can be seen, when the attacker sends a fake frame to the victim, the victim sends back an ACK to the fake MAC address (aa:bb:bb:bb:bb:bb). This experiment confirms that WiFi devices acknowledge  frames without checking their validity. 
Finally, to see if this behavior exists on other WiFi devices, we have repeated this test with a variety of devices (such as laptops, smart thermostats, tablets, smartphones, and access points) with different WiFi chipsets from different vendors, as shown in Table~\ref{tbl:controled-devices}. Note, target devices are connected to a private network and the attacker does not have their secret key. After performing the same experiment as before, we found that all of these devices also respond to fake packets received from a device outside of their network.

\begin{table}[t]
\centering
\begin{tabular}{|l|l|l|}
     \hline
     Device & WiFi module & Standard  \\
     \hline
     MSI GE62 laptop&Intel AC 3160 & 11ac\\
     Ecobee3 thermostat& Atheros & 11n\\
     Surface Pro 2017& Marvel 88W8897 & 11ac\\
     Samsung Galaxy S8 & Murata KM5D18098 & 11ac\\
     Google Wifi AP &  Qualcomm IPQ 4019 & 11ac\\
     \hline
\end{tabular}
\caption{List of tested chipsets/devices}
\label{tbl:controled-devices}
\end{table}

\subsection{How widespread is this loophole?}\label{sec:testing}
In the previous section, we examined a few different WiFi devices and showed that they are all responding to fake frames from unauthorized devices. Here, we examine thousands of devices to see how widespread this behavior is. In the following, we explain the setup and results of this experiment. 

\vspace{0.05in}
\noindent
\textbf{Setup:} To examine thousands of devices, we mounted a WiFi dongle on the roof of a vehicle and drove around the city to test all nearby devices. For the WiFi dongle, we use the same  Realtek  RTL8812AU  USB  WiFi dongle, and connect it to a Microsoft Surface, running Ubuntu 18.04. We develop a multi-threaded program using the Scapy library~\cite{scapy} to discover nearby devices, send fake 802.11 frames to the discovered devices, and verify that target devices respond to our fake frames. Specifically, our implementation contains three threads. The first thread discovers nearby devices by sniffing WiFi traffic and adding the MAC address of unseen devices to a target list. The second thread sends fake 802.11 frames to the list of target devices. Finally, the third thread checks to verify that target devices respond with an ACK.


\begin{table}[t]
    \centering
    \begin{tabular}{|l|c||l|c|}
        \hline
        \multicolumn{2}{|c||}{WiFi Client Device} & \multicolumn{2}{|c|}{WiFi Access Point}\\
        \hline
        Vendor & \# devices &   Vendor & \# devices  \\
        \hline
        Apple&  143	&	Hitron &  723 \\
        Google&  102	&	Sagemcom &  601 \\
        Intel&  66	&	Technicolor&  410 \\
        Hitron &  65	&	eero &  195 \\
        HP &  63	&	Extreme N. &  188 \\
        Samsung&  56	&	Cisco &  156 \\
        Espressif&  47	&	HP &  104 \\
        Hon Hai&  46	&	TP-LINK &  101 \\
        Amazon &  41	&	Google &  80 \\
        Sagemcom &  38	&	D-Link  &  75 \\
        Liteon &  33	&	NETGEAR &  69 \\
        AzureWave &  30	&	ASUSTek  &  51 \\
        Sonos &  30	&	Aruba &  46 \\
        Nest Labs &  27	&	SmartRG, &  44 \\
        Murata  &  24	&	Ubiquiti N.&  35 \\
        Belkin &  20	&	Zebra &  35 \\
        TP-LINK  &  20	&	Pegatron &  28 \\
        Cisco&  16	&	Belkin  &  25 \\
        ecobee &  13	&	Mitsumi &  25 \\
        Microsoft &  13	&	Apple &  19 \\
        Others & 630	&	Others & 789 \\
        \hline\hline	
        Total & 1523	&	Total & 3805  \\
        \hline
    \end{tabular}
    \caption{List of WiFi devices and APs that respond to our fake 802.11 frames.}
    \label{tbl:uncontroled-devices}
\end{table}

\vspace{0.05in}
\noindent
\textbf{Results}: We perform this experiment for one hour while driving around the city. In total, we discovered 5,328 WiFi nodes from 186 vendors. The list includes 1,523 different WiFi client devices from 147 vendors and 3,805 access points from 94 vendors. Table~\ref{tbl:uncontroled-devices} shows the top 20 vendors for WiFi devices and WiFi access points in terms of the number of devices discovered in our experiment. The list includes devices from major smartphone manufacturers (such as Apple, Google, and Samsung) and major IoT vendors (such as Nest, Google, Amazon, and Ecobee). We found that all 5,328 WiFi Access Points and devices responded to our fake 802.11 frames with an acknowledgment, and hence we infer that most probably all of today's WiFi devices and access points respond to fake frames when they should not.

\subsection{Can this loophole be fixed?}\label{sec:cannot-be-fixed}
So far, we have demonstrated that all existing WiFi devices respond to fake packets received from unauthorized WiFi devices outside of their network. Now, the next question is why this behavior exists, and if it can be prevented in future WiFi chipsets.

In a WiFi device, when the physical layer receives a frame, it checks the correctness of the frame using error-checking mechanisms (such as CRC) and transmits an ACK if the frame has no error. However, checking the validity of the content of a frame is performed by the MAC and higher layers. Unfortunately, this separation of responsibilities and the fact that the physical layer does not coordinate with higher layers about sending ACKs seem to be the root cause of the behavior. In particular, we have observed that when some access points receive fake frames, they start sending \emph{deauthentication frames} to the attacker, requesting it to leave the network. 
These access points detect the attacker as a ``malfunctioning'' device and that is why they send deauthentication frames. Surprisingly, although the access points have detected that they are receiving fake frames from a ``malfunctioning'' device, we found that they still acknowledge the fake frames.

\begin{figure}[t!]
    \centering
    \includegraphics[width=\columnwidth]{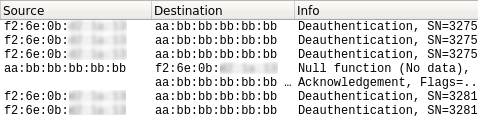}
    \caption{The attacked access point detects that something strange is happening, however it still ACKs fake frames}
    \label{fig:deauth-rts-cts}
\end{figure}

An example traffic that demonstrates this behavior is shown in Figure~\ref{fig:deauth-rts-cts}. As can be seen, although the access point has already sent three deauthentication frames to the attacker, it still acknowledges the attacker's fake frame. We then manually blocked the attacker's fake MAC address on the access point. Surprisingly, we observed that the AP still acknowledges the fake frames. These observations verify that sending ACK frames happens automatically in the physical layer without any communication with higher layers. Therefore, the software running on the access points does not prevent the physical layer from sending ACKs to fake frames.

The next question is why the software running on WiFi devices does not prevent this behavior by verifying if the frame is legitimate before sending an ACK. Unfortunately, this is not possible due to the WiFi standard timing requirements. Specifically, in the IEEE 802.11 standard, upon receiving a frame, an ACK must be transmitted by the end of the Short Interframe Space (SIFS)\footnote{The SIFS is used in the 802.11 standard to give the receiver time to go through different procedures before it is ready to send the ACK.
These procedures include Physical-layer and MAC-layer header processing, creating the waveform for the ACK, and switching the RF circuit from receiving to transmitting mode.
} interval which is 10~$\mu$s and 16 $\mu$s for the 2.4 GHz and 5 GHz bands, respectively.
If the transmitter does not receive an ACK by the end of SIFS, it assumes that the frame has
been lost and retransmits the frame. Therefore, the WiFi device nefeds to verify the validity of the received frame in less than 10 $\mu s$. This verification must be done by decoding the frame using the secret shared key. Unfortunately, decoding a frame in such a short period is not possible. In particular, past work has shown that the time required to decode a frame is between 200 to 700 $\mu s$ when the WPA2 security protocol is used~\cite{decoding-time-1, decoding-time-2, decoding-time-3}. This processing time is orders of magnitude longer than SIFS. Hence, existing devices cannot verify the validity of the frame before sending the ACK, and they acknowledge a frame as long as it passes the error detection check. One potential approach to solve this loophole is to implement the security decoder in WiFi hardware instead of software to significantly speed up its delay. Although this may solve the problem in future WiFi chipsets, it will not fix the problem in billions of WiFi chipsets which are already deployed.

%% file: keep-awake.tex
\section{WiFi Stays Awake When It Should Not}
\begin{figure}[!ht]
    \centering
    \begin{subfigure}[b]{0.24\textwidth}
        \centering 
        \includegraphics[width=\textwidth]{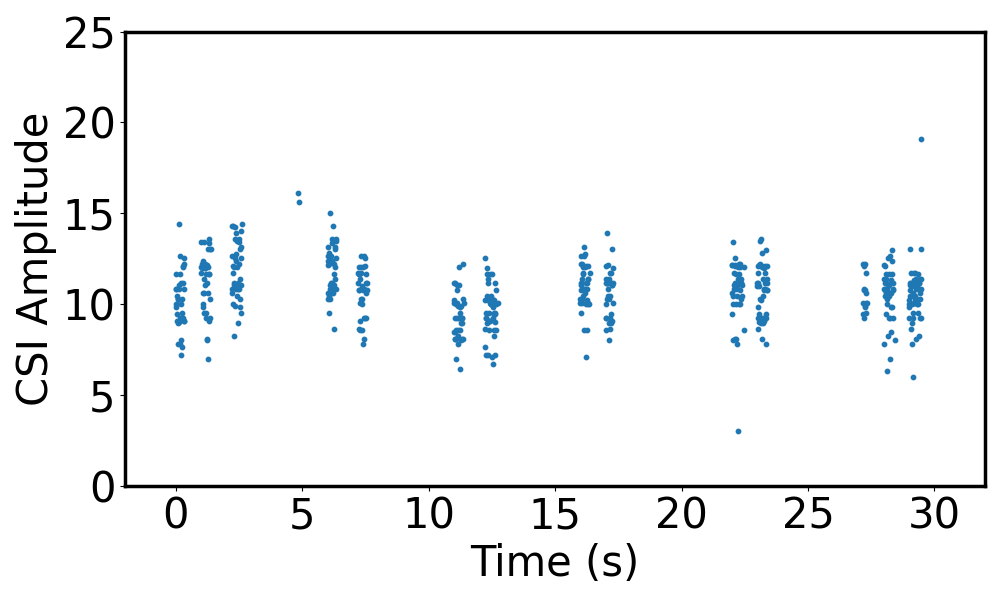}
        \caption{Without fake beacon frames}
        \label{fig:dis}
    \end{subfigure}
    \begin{subfigure}[b]{0.24\textwidth}
        \centering
        \includegraphics[width=\textwidth]{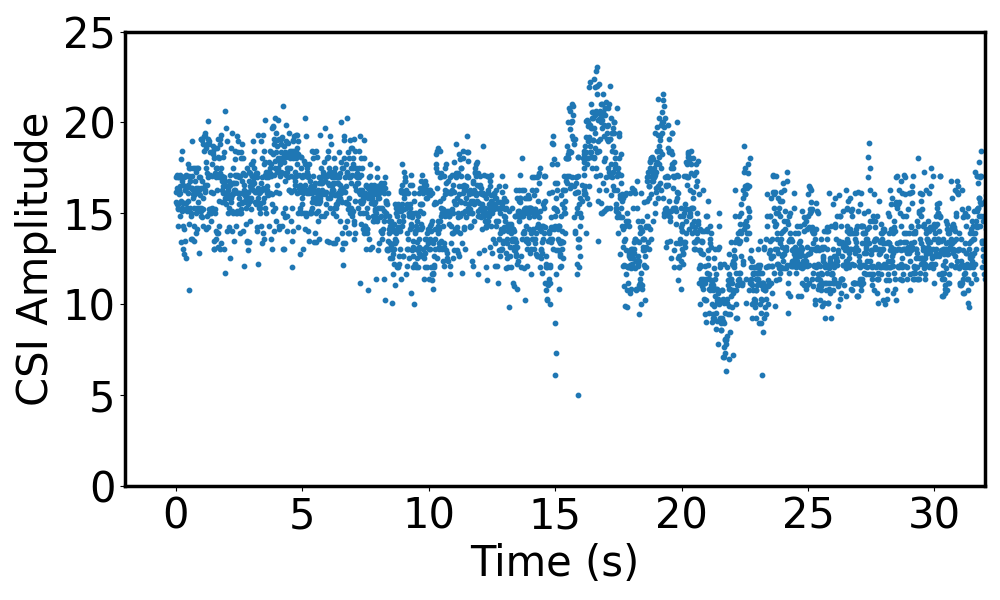}
        \caption{With fake beacon frames}
        \label{fig:cont}
    \end{subfigure}
    \caption{The CSI amplitude of ACKs responded by the target device when an attacker sends back-to-back fake packets to it in two scenarios. (a) In this scenario, the attacker is not using fake beacon frames. Therefore, the target device goes to sleep mode frequently and does not respond to fake packets. (b) In this scenario, the attacker infrequently sends fake beacon frames to keep the target device awake all the time.}
    \label{fig:time-comp}
\end{figure}

We have also found a loophole that allows an unauthorized device to keep a WiFi device awake all the time. One may think that a WiFi device can be kept awake by just sending fake back-to-back packets to it and forcing it to transmit acknowledgment. However, this approach does not work. Most WiFi radios go to sleep mode to save energy during inactive states such as screen lock, during which the attacker is not able to keep them awake by sending back-to-back packets. Figure~\ref{fig:dis} show the results of an experiment where the attacker is continuously transmitting fake packets to a WiFi device. In this figure, we plot the amplitude of CSI over time for the ACK packets received from the WiFi device. As can be seen, the responses are sparse and discontinued even when the attacker sends back-to-back packets to the WiFi device. This is because the WiFi device goes to sleep mode frequently. However, we have found a loophole in the power saving mechanism of WiFi devices which can be used by an unauthorized device to keep any WiFi device awake all the time.

\subsection{How does WiFi power saving mechanism work?}
Wireless tranceivers are very power-hungry. 
Therefore, WiFi radios spend most of the time in the sleep mode to save power. When a WiFi radio is in sleep mode, it cannot send or receive WiFi packets. To avoid missing any incoming packets, when a WiFi device wants to enter the sleep mode it notifies the WiFi access point so that the access point buffers any incoming packets for this device. WiFi devices, however, wake up periodically to receive beacon frames to find out if packets are waiting for them. In particular,  WiFi access points broadcast beacon frames periodically which includes a Traffic Indication Map (TIM) field that indicates which devices have buffered packets on the access point. For example, if the association ID of a WiFi device is $x$, then the $(x+1)^{th}$ bit of TIM is assigned to that device. Finally, when a device is notified that has some buffered packets on the access point, it stays awake and replies with a \textit{Null-function} packet with a power management bit set to "0". In this way, the WiFi device informs the access point it is awake and ready to receive packets.

\subsection{How can one manipulate power saving?}
We have found that an unauthorized device can use the power-saving mechanism of WiFi devices to force them to stay awake. In particular, an attacker can pretend to be the access point and broadcasts fake beacon frames indicating that the WiFi device has buffered traffic, forcing them to stay awake. However, this requires the attacker to know the MAC address and the SSID of the network's access point, as well as the association ID and MAC address of the targeted device so that it can set the correct bit in TIM. 
The access point MAC address and SSID can be easily discovered by sniffing the WiFi traffic using software such as Wireshark since the MAC address is never encrypted and all nodes send packets to the access point. 
Note that MAC randomization does not cause any problem for this process because the attacker finds the randomized MAC address that is currently being used.
Next, the attacker pretends to be the access point and broadcasts fake beacon frames with TIM set to "0xFF", indicating all client devices have buffered traffic. Then, it enters the sniffing mode to sniff for the \textit{Null-function} packets. The null-function packets contain the ID and MAC addresses of all WiFi devices. To avoid keeping all WiFi devices awake, we find that one can send a fake beacon frame as a unicast packet, instead of the usual broadcast beacons. This way only the target device receives the packet and we do not interfere with the operation of other devices. Interestingly, our experiments show that target devices do not care if they receive beacons as broadcast or unicast frames.

\begin{figure}[!t]
    \centering
    \includegraphics[width = 0.8\columnwidth]{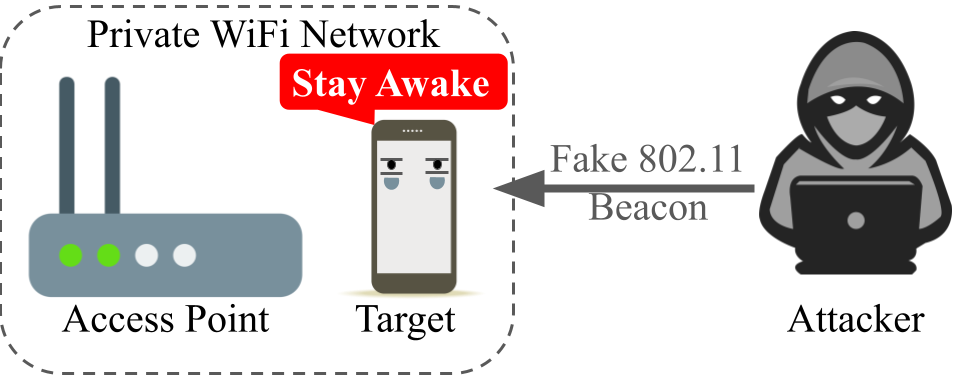}
    \caption{WiFi devices stay awake on hearing a forged beacon frame with TIM flags set up.}
    \label{fig:polite-wifi-beacon}
\end{figure}

To better understand this behavior, we run an experiment where we use two WiFi devices to act as a victim and an attacker, respectively. The attacker sends fake WiFi packets to the victim. We monitor the real traffic between the attacker and the victim's device.

\vspace{0.05in}
\noindent
\textbf{Setup:} Similar to the experiment described in Section~\ref{sec:polite-wifi}, we use an RTL8812AU USB dongle to inject fake packets to a smartphone held by a person who is watching YouTube on the phone. The distance between the smartphone and the user is about 60 cm. The attacking device and the victim are in two separate rooms. The attacker also uses an ESP32 WiFi module to record the Channel State Information (CSI) of received ACKs. 

\vspace{0.05in}
\noindent
\textbf{Result:}
We find that although sending fake beacon frames keeps the target device awake, sending them very frequently will cause WiFi devices to recognize the suspicious attacker's behavior and disconnect from it. Therefore, to keep the WiFi device awake, instead of just sending beacon frames back-to-back, the attacker can continuously transmit normal fake packets to a WiFi device and periodically sends fake beacon frames to keep it awake. Figure~\ref{fig:cont} shows the result of an experiment where the attacker is continuously transmitting fake packets to a WiFi device and periodically sends fake beacon frames. As it can be seen, the target device is continuously awake and responding to fake packets with ACKs. 

%% file: Implicaiton_Sense.tex
\begin{figure*}[th!]
    \centering
    \begin{subfigure}[b]{0.329\textwidth}
        \centering 
        \includegraphics[width=\textwidth]{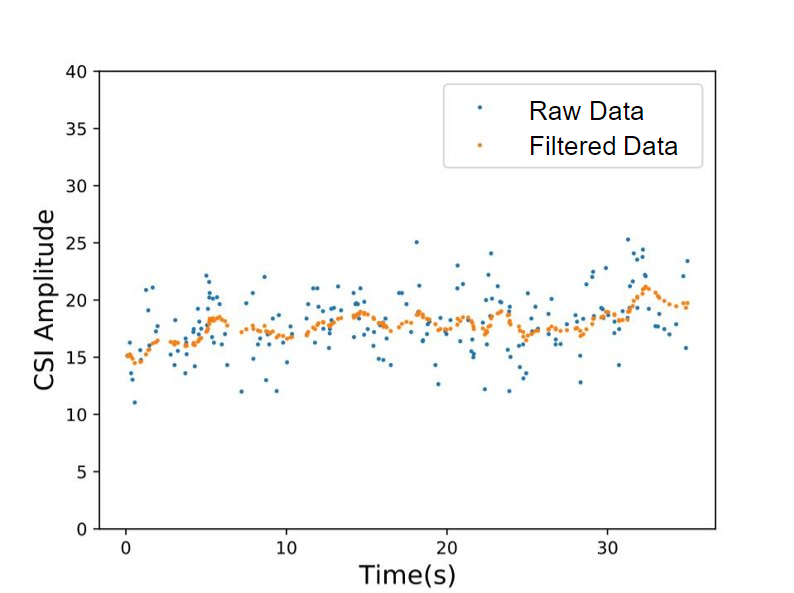}
        \caption{Raw and filtered data before \\ interpolation}
    \end{subfigure}
    \begin{subfigure}[b]{0.329\textwidth}
        \centering
        \includegraphics[width=\textwidth]{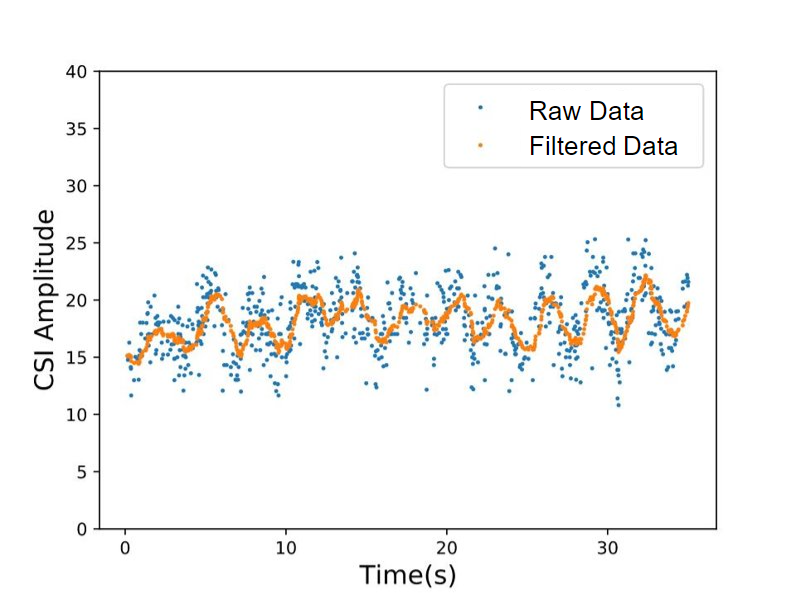}
        \caption{Raw and filtered data after \\ interpolation}
    \end{subfigure}
        \begin{subfigure}[b]{0.328\textwidth}
        \centering
        \includegraphics[width=\textwidth]{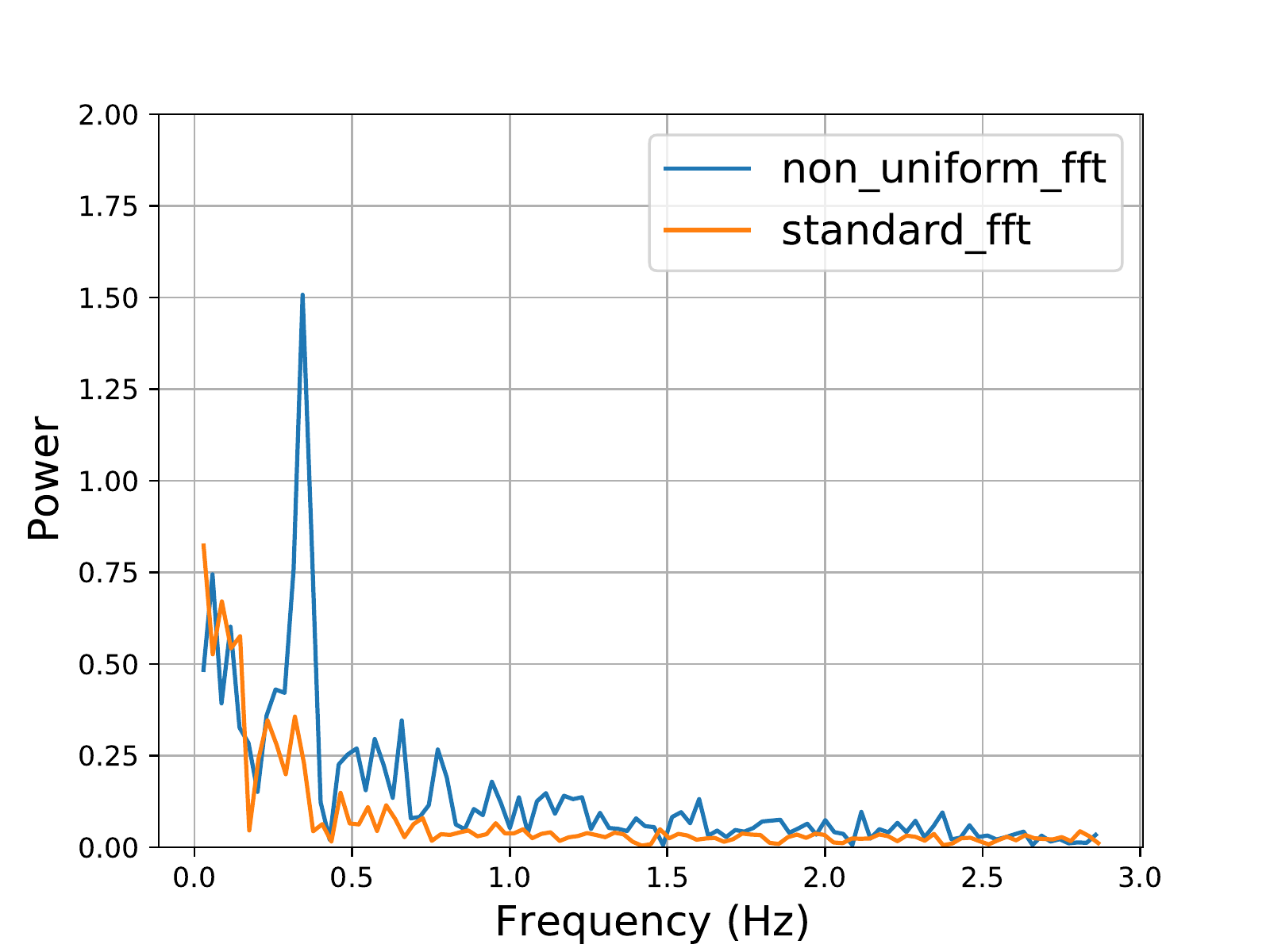}
        \caption{Standard FFT and a non-uniform FFT of Data}
    \end{subfigure}
    \caption{Steps to extract breathing rate from the CSI.}
    \label{fig:process-step}
\end{figure*}

\section{Privacy Implication: WiFi Sensing Attack}\label{sec:implications}

Recently, there has been a significant amount of work on WiFi sensing technologies that use WiFi signals to detect events such as motion, gesture, and breathing rate. In this section, we show how an adversary can combine WiFi sensing techniques with the above loopholes to monitor people's breathing rate whenever she/he wants from outside buildings despite the WiFi network and devices being completely secured. In particular, an adversary can force our WiFi devices to stay awake and continuously transmit WiFi signals. Then she/he can continuously analyze our signals and extract information such as our breathing rate and presents. Note, since most of the time, we are close to a WiFi device (such as a smartwatch, laptop, or tablet), our body will change the amplitude and phase of the signals which can be easily extracted by the adversary.

\subsection{Attack Design, Scenarios and Setup}

\subsubsection{Attack Design}
The attacker sends fake packets to a WiFi device in the target property and pushes it to transmit ACK packets. In particular, since an adult’s normal breathing rate is around 12 -20
times per minute (i.e., 0.2- 0.33Hz), receiving several ACK packets per second is sufficient for the attacker to estimate the breathing rate, without impacting the performance of the target WiFi network. The attacker then takes the Fourier transform of the CSI information of ACK packets to estimate the breathing rate of the person who is nearby the WiFi device. However, due to the random delays of the WiFi random access protocol and the operating
system’s scheduling protocol, the collected data samples are not uniformly spaced in time. Hence, the attacker cannot simply use standard FFT to estimate the breathing rate. Instead, they need to use a non-uniform Fourier transform, and a voting algorithm to extract the breathing rate. The  Non-Uniform Fast Fourier Transform (NUFFT) algorithm \ref{alg:nfft}  used is shown below.

\begin{algorithm}
\SetAlgoLined
\SetKwFunction{Union}{Union}\SetKwFunction{Interpolation}{Interpolation}
\KwData{Time indices $t$, data samples $x$ of length $n$}
\KwResult{Magnitude of each frequency component}
 $d \leftarrow \min_i({t_i - t_{i - 1}}) \quad i = 1, 2, ..., n.$\;
 \For{$i \leftarrow 1$ \KwTo $n - 1$}{
    $interval \leftarrow t[i] - t[i - 1]$\;
  \If{$interval > d$}{
    $count \leftarrow \lfloor interval / d \rfloor$\;
    \textbf{Interpolation}($t$, $x$, $t[i]$, $t[i-1]$, $count$)\;
   }
 }
 \Return \textbf{FFT}($t$, $x$)
 \caption{Non-uniform FFT}
 \label{alg:nfft}
\end{algorithm}

The algorithm first finds the minimum time gap between any two adjacent data points $d$, then linearly interpolates any interval that is larger than the gap with $\lfloor interval / d \rfloor$\$ samples. Finally, it uses a regular FFT algorithm to find the magnitude of each frequency component. A low-pass filter is applied before feeding data to the FFT analysis to reduce noise (not shown in the algorithm).

Figure~\ref {fig:process-step}(a) and ~\ref{fig:process-step}(b) show the amplitude of CSI before and after interpolation, respectively, when the attacker sends 10 packets per second to a WiFi device that is close to the victim. Each figure shows both the original data (in blue) and the filtered data (in orange). Figure~\ref{fig:process-step}(c) shows the frequency spectrum of the same signals when a standard FFT or our non-uniform FFT is applied. A prominent peak at 0.3Hz is shown in the non-uniform FFT spectrum, indicating a breathing rate of 18 bpm.

WiFi CSI gives us the amplitude of 52 subcarriers per packet. We observed that these subcarriers are not equally sensitive to the motion of the chest. Besides, a subcarrier's sensitivity may vary depending on the surrounding environment. For a more reliable attack, the attacker should identify the most sensitive subcarriers over a sampling window. Previously proposed voting mechanisms for coarse-grained motion detection applications \cite{zhu2018tu,  Arshad17,MoSense,WiGest} cannot be directly applied in this situation, as chest motion during respiration is at a much smaller scale. Instead, we developed a soft voting mechanism, where each subcarrier gives a weighted vote to a breathing rate value. The breathing rate that gets the most votes is reported.  Specifically, We first find the power of the highest peak ($P_{peak}$), and then calculate the average power of the rest bins ($P_{ave}$). The exponent of the Peak-to-Average Ratio (PAR): $e^{\frac{f_{peak}}{f_{ave}}}$ is used as the weight of the corresponding subcarrier. In this way, we guarantee the subcarriers with higher SNR have significantly more votes than the rest of the subcarriers.

\subsubsection{Attack Scenarios}
We evaluate the WiFi sensing attack in different scenarios, both indoor and outdoor. In the indoor scenario, the attacker and the target are placed in the same building but on different floors. The height of one floor in the building is around $2.8$ m.  This scenario is similar to when the attacker and the target person are in different units of an apartment or townhouse.
In the outdoor scenario, the attacker is outside the target's house. For the outdoor experiments, We place the attacker in another building which is around 20 m away from the target building. In all of the experiments, the target WiFi devices are placed $0.5$ to $1.4$ m away from the person's body. The person is either watching a movie, typing on a laptop, or surfing the web using his cell phone. During the experiments, other people are walking and living normally in the house. Finally, we run the attack and compare the estimated breathing rate with the ground truth.
To obtain the ground truth, we record the target person's breathing sound by attaching a microphone near his/her mouth~\cite{dafna2015sleep}. We then calculate the FFT on the sound signal to measure the breathing frequency. Note that the attack does not need this information and this is just to obtain the ground truth in our experiments.

\subsubsection{Attacker Setup}
\noindent\textbf{Hardware Setup:}
The attacker uses a Linksys AE6000 WiFi card and an ESP32 WiFi module~\cite{esp32} as the attacking device. Both devices are connected to a ThinkPad laptop via USB. The Linksys AE6000 is used to send fake packets and the ESP32 WiFi module is used to receive acknowledgments (ACK) and extract CSI. Although we use two different devices for sending and receiving, one can simply use an ESP32 WiFi module for both purposes. The use of two separate modules gave us more flexibility in running many experiments. As for the target device, we use a One Plus 8T smartphone without any software or hardware modifications. We have also tested our attack on an unmodified Lenovo laptop, a Microsoft Surface Pro 4 laptop, and a USB WiFi card as the target device and we obtained similar results. It is worth mentioning that any WiFi device can be a target without any software or hardware modification. 

\vspace{0.05in}
\noindent\textbf{Software Setup:}
We have implemented the CSI collecting script on the ESP32 WiFi module, and the breathing rate estimation algorithm on the laptop. The collected CSI data is fed to the algorithm which produces the breathing rate estimation values in real-time.  To process this data in real time, a sliding window (buffer) is used. The size of the window is $30$ s and the stride step is $1$ s. 
$30$ seconds is a large enough window for estimating a stable breathing rate value. Note that an adult breathes around 6 times during such a window. The window is a queue of data points, and it updates every second by including $1$ second of new data points to its head and removing $1$ second of old data points from its tail. The breathing rate estimation runs the analysis algorithm on the data points inside the window whenever it is updated. The window slides once per second. Hence, our software reports an estimation of breathing rate every second. Note that there is a $30$-second delay at the beginning since the window needs to be filled first.

\begin{figure*}[t]
    \centering
    
    \setkeys{Gin}{width=\linewidth}
    \begin{tabularx}{\linewidth}{XXX}
    
    \includegraphics[width=0.32\textwidth]{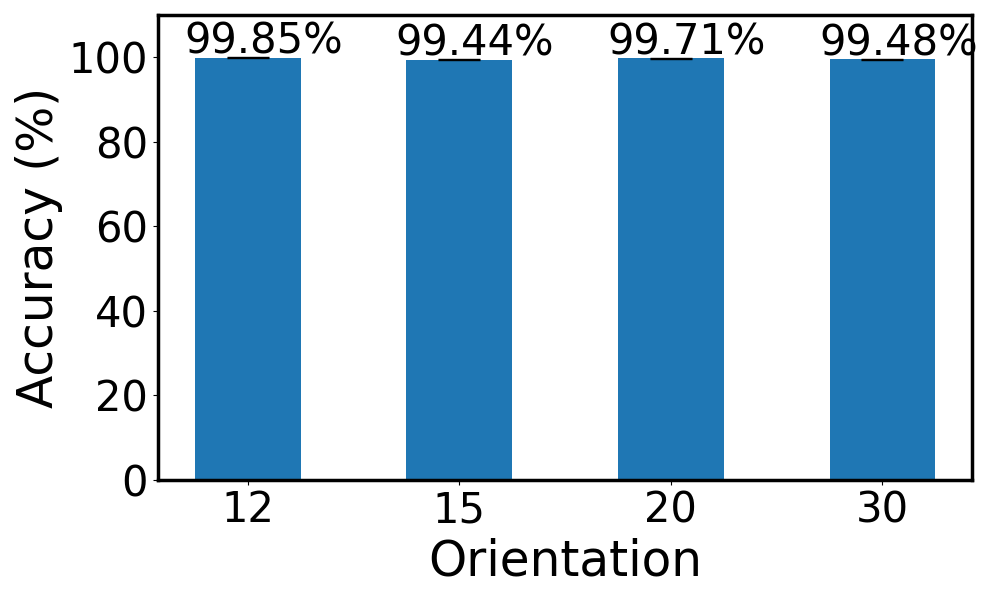}
    \caption{The average accuracy of the attack in estimating the target person's breathing rate when he attacker and target device are in the same building.}
    \label{fig:breath_accuracy_bar}
    &
    \includegraphics[width=0.32\textwidth]{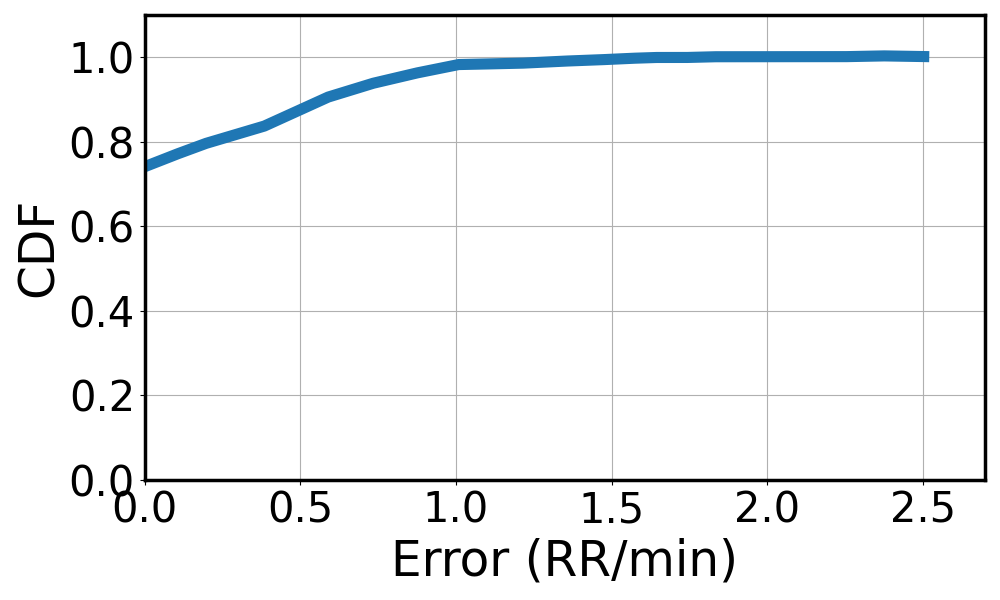}
    \caption{The CDF of the error in estimating the target person's breathing rate when he attacker and target device are in the same building (different floor).}
    \label{fig:breath_accuracy_cdf1}
    &
    \includegraphics[width=0.32\textwidth]{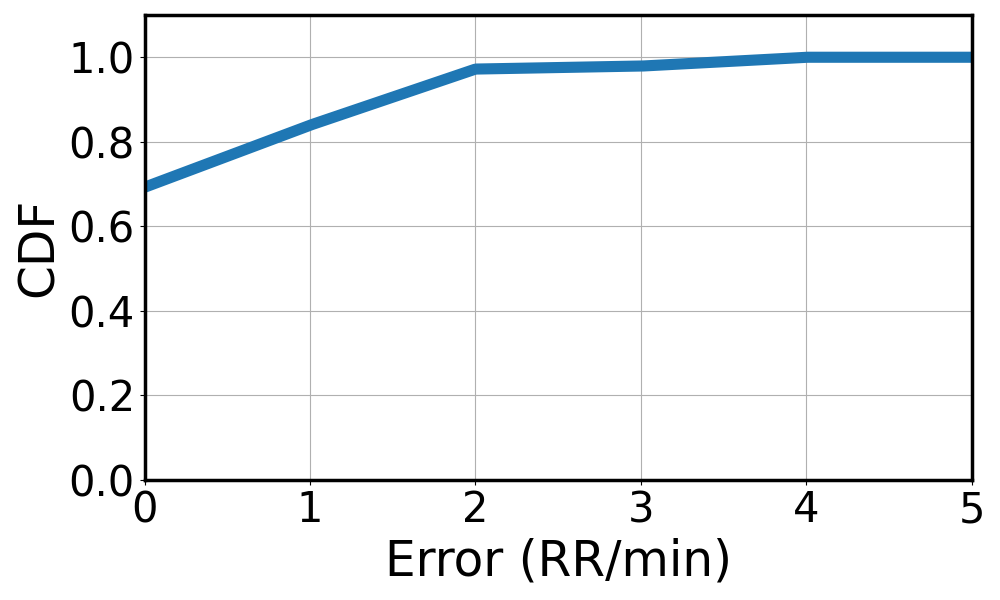}
    \caption{The CDF of the error in estimating the target person's breathing rate when he attacker and target device are in different buildings (20m away)}
    \label{fig:breath_accuracy_cdf2}
 \end{tabularx}    
\end{figure*}

\subsection{Results}
We evaluate the effectiveness of the attack in different scenarios such as when the attacker and the target are in the same building or different buildings.
\subsubsection{Accuracy in Detecting Breathing Rate}
\noindent\textbf{Same Building Scenario:} First, we evaluate the accuracy of the attack by estimating the breathing rate in an indoor scenario where the target device and attacker are in the same building. We evaluate the accuracy when the target's breathing rate is 12, 15, 20, and 30 breaths per minute.  Note, that the normal breathing rate for an adult is 12-20 breaths per minute while resting, and higher when exercising. In this experiment, the user is watching a video. To make sure the target person's breathing rate is close to our desired numbers, we place a timer in front of the person, where they can adjust their breathing rate accordingly. 
This is just to better control the breathing rate during the experiment and is not a requirement nor an assumption in this attack.
We run each experiment for two minutes. During this time, we collect the estimated breathing rate from both ground truth and the attack for different locations of the target device. Figure \ref{fig:breath_accuracy_bar} shows the average accuracy in estimating breathing rate across all experiments. The accuracy is calculated as the ratio of the estimated breathing rate reported by the attack over the ground truth breathing rate. The figure shows that the accuracy of estimating the breathing rate is over $99$\% in all scenarios.  Finally, Figure \ref{fig:breath_accuracy_cdf1} plots the Cumulative Distribution Function (CDF) of the error in  detecting breathing rate for over $2400$ measurements. The figure shows that $78\%$ of the estimated results have no error. The figure also shows that $99\%$ of measurements have less than one breath per minute error which is negligible.

\vspace{0.05in}\noindent\textbf{Different Building Scenario:}
So far, we have evaluated our attack where the target and the attacker are in different rooms or floors of the same building. Here we push this further and examine whether our attack works if the attacker and the target person are in a different building.  We place the target device in a building on a university campus on a weekday with people around. A person is sitting around 0.5 m away from the device.  We then place the attacker in another building which is around 20 m away from the target building.  Similar to the previous experiment, we run the attack and compare the estimated breathing rate with the ground truth. Figure \ref{fig:breath_accuracy_cdf2} shows the CDF of error for 180 measurements in this experiment. Our results show that the attacker successfully estimates the breathing rate. Note, that the reason that the attack works even in such a challenging scenario with other people being around is two-fold. First, using an FFT helps to filter out the effect of most non-periodic movements and focuses on periodic movements and patterns. Second, wireless channels are more sensitive to changes as we get closer to the transmitter \cite{abedi2020witag,dehbashi2021verification}, and since in these scenarios, the target person is very close to the target device, their breathing motion has a higher impact on the CSI signal compared to the other mobility in the environment. 

\subsubsection{Human Presence Detection}
\begin{figure}[!t]
    \centering
    \includegraphics[width=0.45\textwidth]{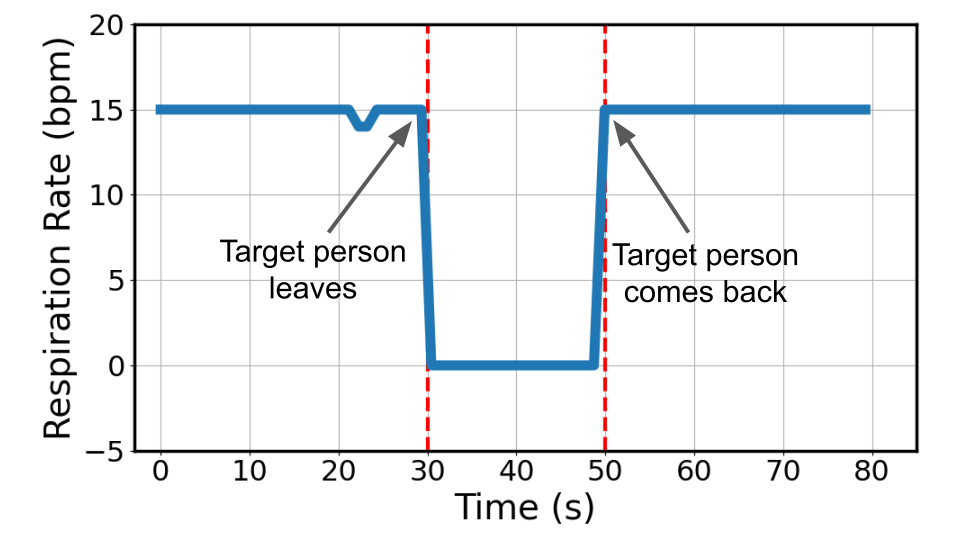}
    \caption{The efficacy of estimating the breathing rate when there is no target near the WiFi device.}
    \label{fig:absence}
\end{figure}

We next evaluate the efficacy of detecting whether there is a target person near the WiFi device or not. In this experiment, the target phone is placed on a desk and the person stays around the device for $30$ seconds, then walks away from the device, and then comes back near the device. Note, in our algorithm, when there is no majority vote during the voting phase, we return $-1$ to indicate no breathing detected. Figure~\ref{fig:absence} shows the results of this experiment. As illustrated in the figure, we can correctly detect the breathing rate when a person is near the device. In other words, the algorithm can detect if there is no one near the target device and refrain from reporting a random value.

\subsubsection{Effect of Distance and Orientation}
Next, we evaluate the effectiveness of the attack for different orientations of the device with respect to the person. We also evaluate its performance for different distances between the target device and the target person.

\begin{figure}[t]
    \centering
    \begin{subfigure}[b]{0.32\textwidth}
        \centering 
        \includegraphics[width=\textwidth]{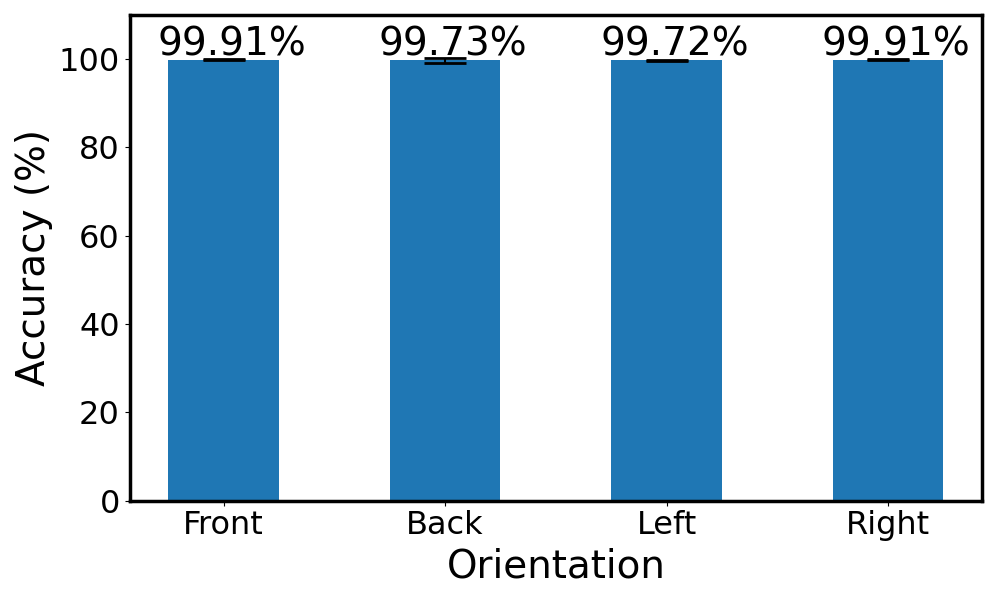}
         \caption{various orientations}
        \label{fig:orientation_vs_accuracy}
    \end{subfigure}
    \hfill
    \begin{subfigure}[b]{0.32\textwidth}
        \centering
        \includegraphics[width=\textwidth]{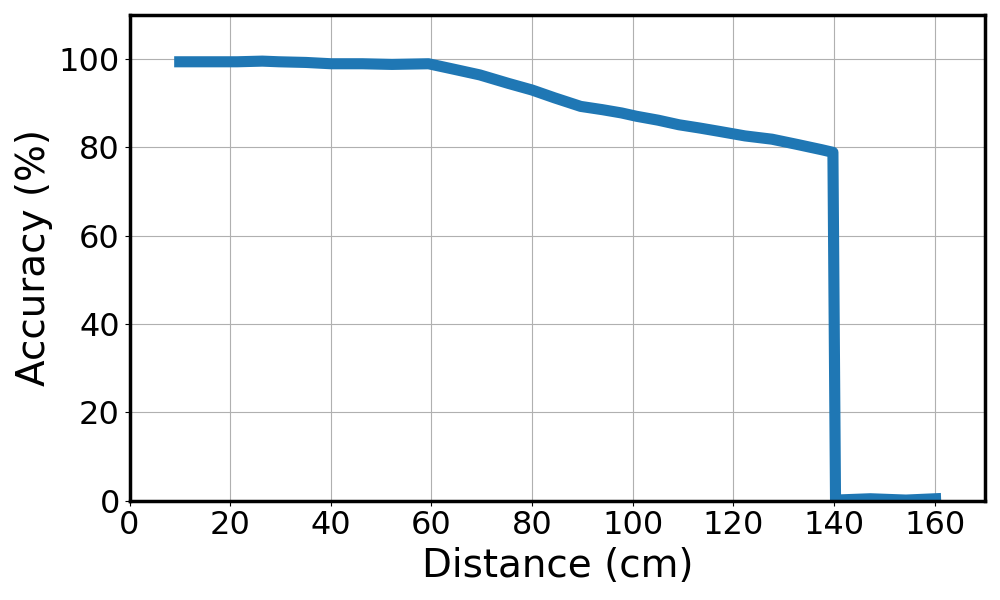}
        \caption{different distances.}
        \label{fig:dist_vs_accuracy}
    \end{subfigure}
    \caption{effectiveness of the attack for different orientation and distance of the targeted WiFi device respect to the person.}
\end{figure}

\vspace{0.05in}
\noindent\textbf{Orientation:}
We evaluate the effect of orientation of the target person with respect to the target device (laptop). We run the same attack as before for different orientations (i.e. sitting in front, back, left, and right side of a laptop). The user is 0.5m away from the target device in all cases. Figure~\ref{fig:orientation_vs_accuracy} shows the result of this experiment. Each bar shows the average accuracy for 90 measurements. Our result shows that regardless of the orientation of the person with respect to the device, the attack is effective and detects the breathing rate of the person accurately. In particular, even when the person was behind the target device, the attack still detects the breathing rate with 99\% accuracy.

\vspace{0.05in}
\noindent
\textbf{Distance:}
Here,  we are interested to find out what the maximum distance between the target device and the person can be while the attacker still detects the person's breathing rate. To do so, we place the attacker device and the target device 5 meters apart in two different rooms with a wall in between. We then run different experiments in which the target person stays at different distances from the target device. In each experiment, we measure the breathing rate for two minutes and calculate the average breathing rate over this time. Finally, we compare the estimated breathing rate to the ground truth and calculate the accuracy as mentioned before.

Figure~\ref{fig:dist_vs_accuracy} shows the results of this experiment.  The accuracy is over $99\%$ when the distance between the target device and the target person is less than $60$ cm. Note, in reality, people have their laptops or cellphone very close to themselves most of the time, and $60$ cm is representative of these situations. The accuracy drops as we increase the distance. However, even when the device is at 1.4 m from the person's body, the attack can still estimate the breathing rate with $80\%$ accuracy. Note, this is the accuracy in finding the absolute breathing rate and the change in the breathing rate can be detected with much higher accuracy. Finally, the figure shows that the accuracy suddenly drops to zero for a distance beyond 1.4 m. This is due to the fact that at that distance the power of the peak at the output of the FFT goes below the noise floor, and hence, the peak is not detectable.

\subsubsection{Effect of Multiple People}

\begin{figure*}[!t]
    \centering
    \begin{subfigure}[b]{0.19\textwidth}
        \centering 
        \includegraphics[width=\textwidth]{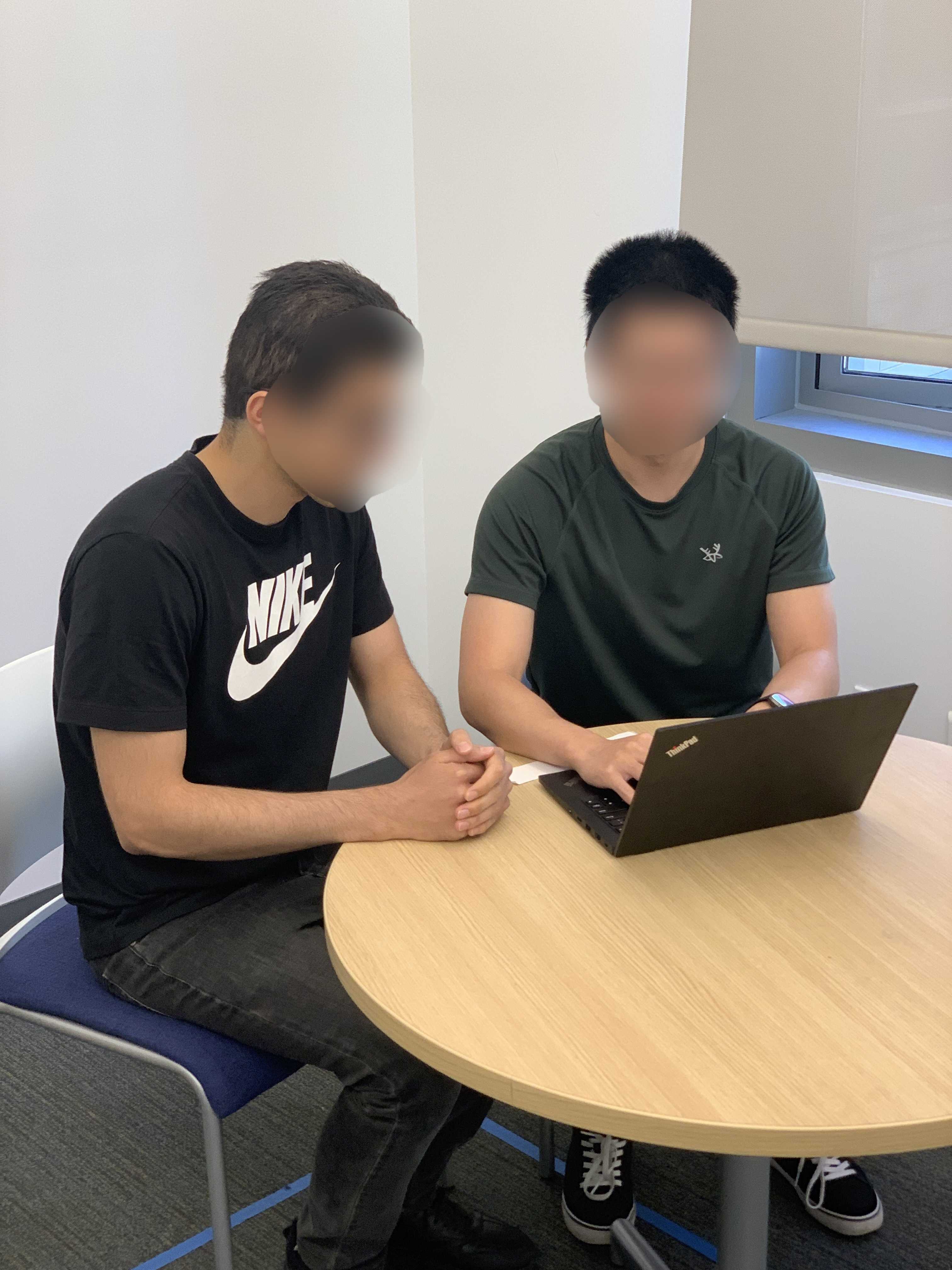}
        \caption{Scenario 1}
        \label{fig:scene1}
    \end{subfigure}
    \hfill
    \begin{subfigure}[b]{0.34\textwidth}
        \centering
        \includegraphics[width=\textwidth]{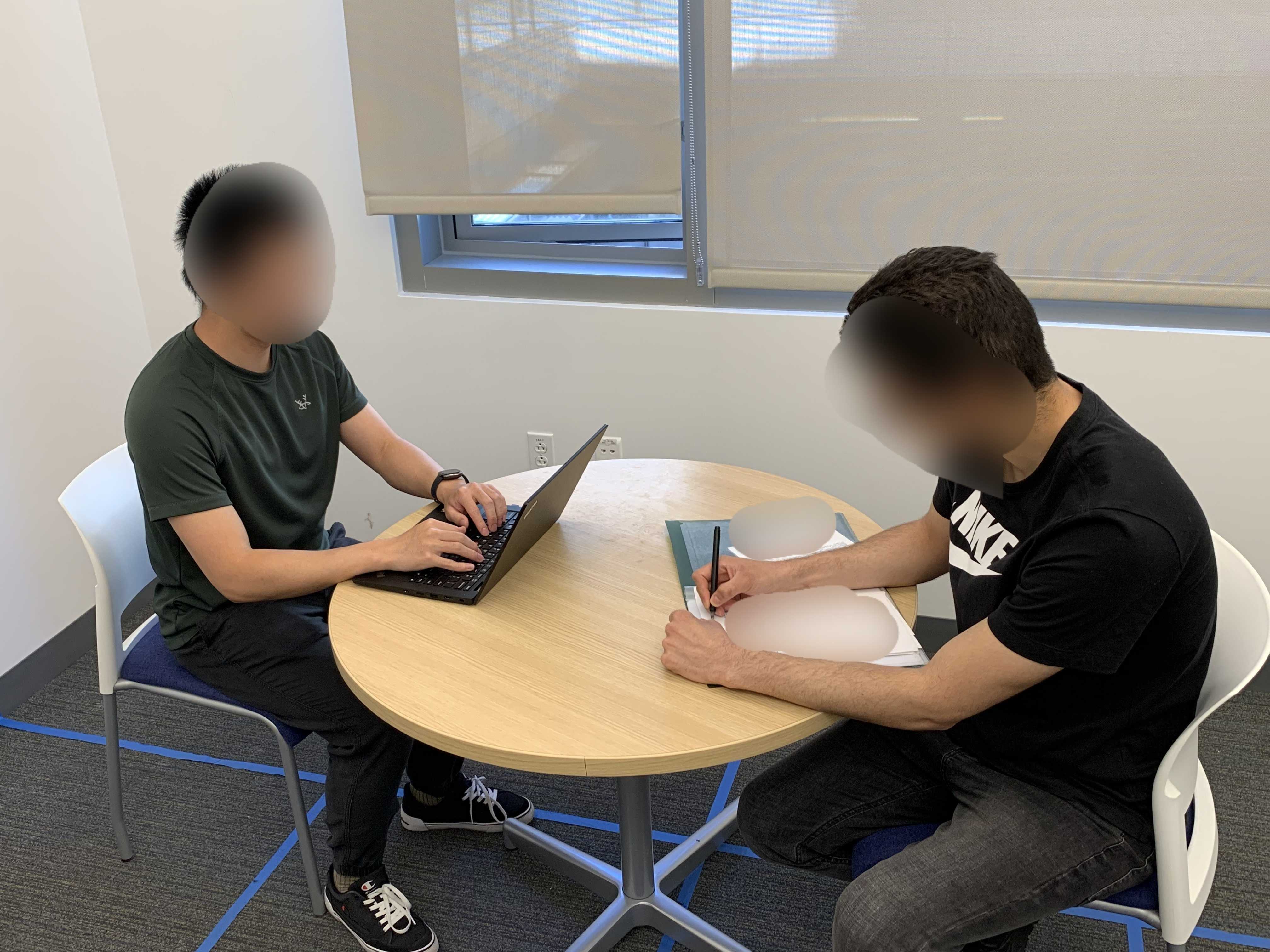}
        \caption{Scenario 2}
        \label{fig:scene2}
    \end{subfigure}
    \begin{subfigure}[b]{0.4\textwidth}
        \centering
        \includegraphics[width=\textwidth]{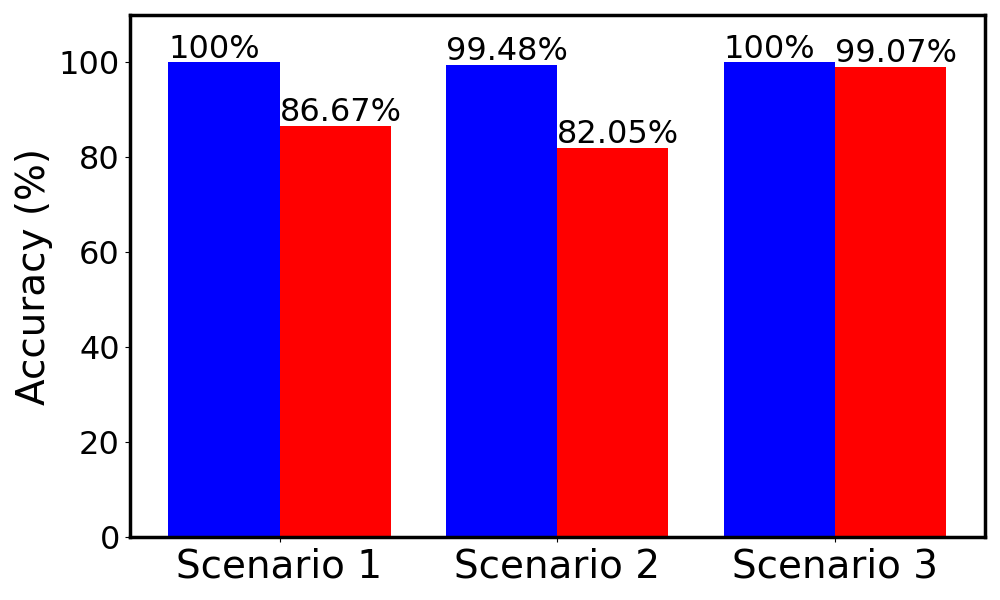}
        \vspace{-18pt}
        \caption{Breathing Rate Estimation of two persons}
        \label{fig:scenarios_vs_accuracy}
    \end{subfigure}
    \caption{Accuracy under three different scenarios: Scenario 1: two people sit side-by-side in front of the target device; Scenario 2: one person sits in front of the target device, the other one sits behind the target device; Scenario 3: two people sit in front of two target devices, respectively. Attacker attacks one by one.}
    \label{fig:two-scenes}
\end{figure*}
Last, we evaluate if the attack can be used to detect the breathing rate of multiple people simultaneously. We test our attack in three different scenarios. In the first scenario, two people are near the laptop while one is working on the laptop and the other is just sitting next to him, as shown in Figure \ref{fig:scene1}. The attacker targets the laptop and tries to estimate their breathing rate. Note, that the attacker has no prior information about how many people are next to the laptop. In the second scenario, we repeat the same experiment as the first scenario except that the second person is sitting behind the laptop, as shown in Figure \ref{fig:scene2}. In the third scenario, there are two people in the same space but each person is next to a different device. The attacker targets the laptops and tries to estimate their breathing rates. In these experiments, the target device is 0.5-0.7 m away from the person. 

Figure \ref{fig:scenarios_vs_accuracy} shows the results for this evaluation. The blue bars show the result for the first person who is working on the laptop, and the red bars show the results for the second person. Our results show that the attack effectively detects the breathing rate of both people regardless of their orientation. However, the accuracy in detecting the breathing rate for the second person is a bit lower than the first person for the first and second scenarios. This is because the second person's distance to the target device is slightly more and hence the accuracy has decreased.

%% file: Implicaiton_Battery.tex
\section{Security Implication: Battery Drain Attack}
 In this section, we show how an adversary can drain the battery of our WiFi devices by using the above loopholes and forcing our WiFi devices to stay awake and continuously transmit WiFi signals.

\subsection{Attack Design and Setup}

\subsubsection{Attack Design}
\label{BatteryAttackDesign}
The attacker forces the target device to stay awake and continuously transmit WiFi packets by sending it back-to-back fake frames and some periodic fake beacons. However, to maximize the amount of time the target device spends transmitting, we study a few different types of fake query packets that the attacker can send. 
Note, that the power consumption of transmission is typically higher than that of reception.\footnote{For example, ESP8266~\cite{esp8266} and ESP32~\cite{esp32} WiFi modules draw 50 and 100 mA when receiving while they draw 170 and 240 mA when transmitting.  These low-power WiFi modules are very popular for IoT devices~\cite{abedi2019wi}.} Hence, to maximize the battery drain, we want to send a short query packet and receive a long response.

\begin{table}[h]
    \centering
    \begin{tabular}{|l|l|l|l|}
        \hline
         Query & Query size & Response & Response size \\
         \hline
         Null    &   28 bytes & ACK   &   14 bytes\\
         RTS            &  20 bytes & CTS   &  14 bytes\\
         BAR     &  24 bytes & BA &  32 bytes \\
         \hline
    \end{tabular}
    \caption{Different types of fake queries and their  responses. Note, Null is a data packet without any payload. BAR and BA stand for Block ACK Request, and Block ACK, respectivly.}
    \vspace{-10pt}
    \label{tbl:queries}
\end{table}

Table~\ref{tbl:queries} lists some query packets and their corresponding responses. The best choice for a query packet is Block ACK requests since the target will respond with a Block ACK that is larger than other query responses. Another important factor to consider for maximizing the battery drain is the bitrate.  When the bitrate of the query packet increases, the bitrate of the response will also increase as specified in the IEEE 802.11 standard. Hence, at first glance, it may seem that to maximize the battery drain, the attacker must use the fastest bitrate possible to transmit query packets, forcing the target device to transmit as many responses as possible. However, it turns out that this is not the case. The power consumption depends mostly on the amount of time the target device spends transmitting packets. Hence, when a higher rate is used for the query and response packets, the total time the target spends on transmission does not increase. In fact, the total time spent transmitting decreases mainly due to overheads such as channel sensing and backoffs. For example, if we increase the bitrate by 6 times (i.e., from 1 Mbps to 6 Mbps), the number of packets will increase by only 3.3 times. As a result, to maximize the transmission time of the target device, the attacker should use the lowest rate (i.e., 1 Mbps) for the query packet.

\subsubsection{Attack Setup}
\vspace{0.05in}
\noindent

\noindent\textbf{Attacking device:}
Any WiFi card capable of packet injection can be used as the attacking device. We use a USB WiFi card connected to a laptop running Ubuntu 20.04. The WiFi card has an RTL8812AU chipset~\cite{rtl8812au} that supports IEEE 802.11 a/b/g/n/ac standards. We have installed the aircrack-ng/rtl8812au driver~\cite{aircrack-ng} for this card which enables robust packet injection. We utilize the Scapy~\cite{scapy} library to inject fake WiFi packets to the target device. 
Scapy allows defining customized packets and multiple options for packet injection. 
Since we need to inject many packets in this attack, we use the \textit{sendpfast} function to inject packets at high rates. \textit{sendpfast} relies on \textit{tcpreplay}~\cite{tcpreplay} for high performance packet injection. 






\noindent
\textbf{Target device:}
Any WiFi-based IoT device can be used as a target.  We choose Amazon Ring Spotlight Cam Battery HD Security Camera~\cite{ringcamera}
for our battery drain experiments. The camera is powered by a custom 6040 mAh lithium-ion battery. 
The battery life of this camera is estimated to be between 6 and 12 months under normal usage~\cite{xyz, bat2}.
We leave the camera settings to their defaults which means most power-consuming options are turned off. This assures that our measurements will be an upper bound on the battery life and hence the attack might drain the battery much faster in the real world.
Authors in \cite{vanhoef2020protecting} pointed out the possibility of a battery draining attack by forging beacon frames. However, they did not provide any evaluations to test this idea. Moreover, we show how sending fake packets in addition to fake beacon frames can significantly increase the power consumption on the victim device.

\subsection{Results}
We evaluate the effectiveness of the battery drain attack in terms of range and using different payload configuration. 

\begin{figure}[t]
    \centering
    \begin{subfigure}[b]{0.35\textwidth}
    \includegraphics[width=1\columnwidth]{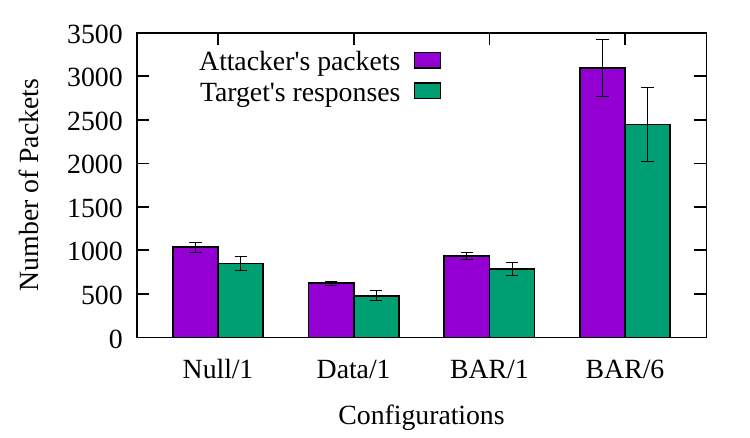}
    \caption{}
    \end{subfigure}
    \begin{subfigure}[b]{0.35\textwidth}
    \includegraphics[width=1\columnwidth]{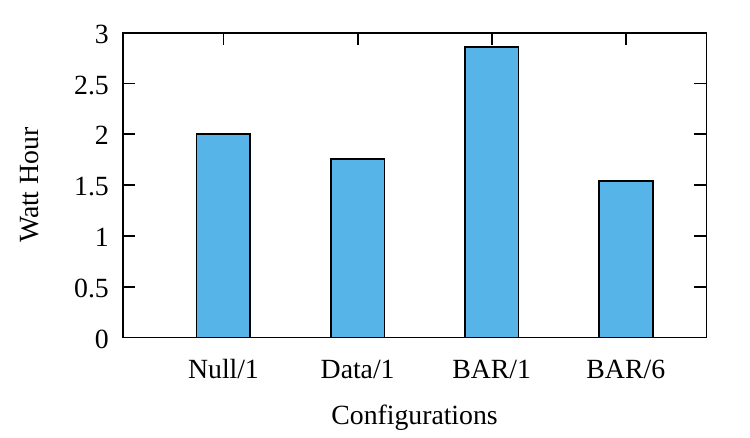}
    \caption{}
    \end{subfigure}
    \caption{The figure shows (a) Average number of packets sent to and received from the target device. (b) Energy consumption in Watt Hour measured under different configurations (i.e. packet type / bitrate (Mbps) }
    \label{fig:batteryDrain}
\end{figure}

\begin{figure*}[t]
    \centering
    \includegraphics[width=0.8\textwidth]{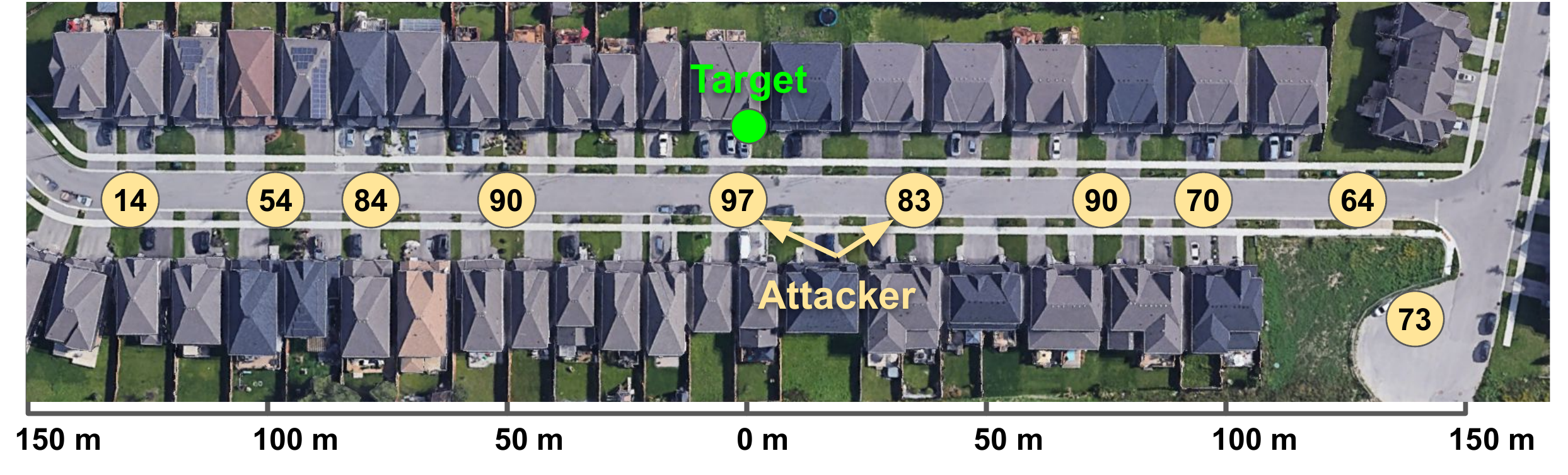}
    \vspace{-10pt}
    \caption{Percentage of attacker's query packets responded by the target device for different attacker's locations.}
    \label{fig:battery-range-map}
\end{figure*}

\begin{table*}[t]
    \centering
    \begin{tabular}{|c|c|c|c|c|}
        \hline
        Battery Type & Voltage (V) & Full Capacity (Wh) & 100\% Drain  (min)  & 25\% Drain (min) \\
        \hline
        CR2032 coin & 3.0  &  0.68  &  14  & 3.5  \\
        AAA    & 1.5  &  1.87  &  39  & 10 \\
        AA     & 1.5  &  4.20  &  90  & 22\\
        \hline
    \end{tabular}
    \caption{The time it takes for the attack to drain different types of batteries}
    \label{tab:my_label}
\end{table*}

\subsubsection{Finding the optimal configuration:}
As discussed in~\ref{BatteryAttackDesign}, sending block ACK requests at the lowest bitrate (i.e., 1 Mbps) should maximize the power consumption of the target device. To verify this, we have conducted a series of experiments with different types of query packets and transmission bitrates. In each experiment, we continuously transmit query packets to the Ring security camera. In all experiments, we start with a fully charged battery and the attacker injects query packets as fast as possible.

Figure~\ref{fig:batteryDrain} (a) shows the maximum number of packets the attacker could transmit to the target device, and the number of responses it receives per second. Figure~\ref{fig:batteryDrain} (b) shows the amount of energy drawn from the battery during one hour of the attack. As expected, sending Block ACK Requests (BAR) drains more energy from the battery since the target device spends more time on transmission than receiving. Moreover, the results verify that although increasing the data rate from 1Mbps to 6Mbps (BAR/1 versus BAR/6) increases the number of responses, it decreases the energy drained. As mentioned before, this is because the total time spent transmitting decreases mainly due to overheads such as channel sensing and backoffs. This result confirms that sending block ACK requests (BAR) with the lowest datarate is the best option to drain the battery of the target device.


\subsubsection{Battery drain with optimal configurations}
We use the best setting which is a block ACK request (BAR) query transmitted at 1~Mbps to fully drain the battery of the Ring security camera. We are able to drain a fully charged battery in 36 hours. Considering the fact that the typical battery life of this camera is 6 to 12 months, our attack reduces the battery life by 120 to 240 times! It is worth mentioning that since a typical user charges the battery every 6-12 months, on average the batteries are at 40-60\%, and therefore it would take much less for our attack to kill the battery. Moreover, the RING security camera is using a very large battery, most security sensors are using smaller batteries. Table~\ref{tab:my_label} shows the amount of time it takes to drain different batteries. For example, it takes less than 40 mins to kill a fully charged AAA battery which is a common battery in many sensors.

\subsubsection{Range of WiFi battery draining attack}
A key factor in the effectiveness of the battery draining attack is how far the attacker can be from the victim's device and still be able to carry on the attack. If the attack can be done from far away, it becomes more threatening. 
To evaluate the range of this attack, we design an experiment in which the attacker transmits packets to the target from different distances and we measure what percentage of the attacker's packets are responded to by the target device.
We use a realistic testbed. The Ring security camera is installed in front of a house, and the attacker is placed in a car, parked at different locations on the street. We test the attack at 10 different locations up to 150 meters away from the target device. Figure~\ref{fig:battery-range-map} 
shows these locations and our setup. Each yellow circle represents each of the locations tested at. The numbers inside the circles show the percentage of the attacker's packets responded to by the camera.
Each number is an average of over 60 one-second measurements.
The closest distance is about 5 meters when we park the car in front of the target house. 
In this location 97\% of the attacker's packets are responded to.
We conducted other experiments within 10 meters of the target (not shown here) and we obtained similar results. Our results show that even within a distance of 100 meters, almost all attacker's packets are responded to by the victim's device. In some locations such as the rightmost circle (at 150 meters away), we could still achieve a reply rate as high as 73\%, confirming our attack works even at that distance. The reason for achieving such a long range is that the attacker transmits at a 1 Mbps bitrate which uses extremely robust modulation and coding rate (i.e. BPSK modulation and a 1/11 coding rate).

%% file: Conclusion.tex
\section{Ethical Considerations}
We discussed our project and experiments with our institutions’ IRB office and they determined that no IRB review nor IRB approval is required. Moreover, the house and WiFi devices used in most experiments are owned and controlled by the authors. Finally, in order to expedite mitigating the attacks presented in this paper, we have started engagements with WiFi access point and chipset manufacturers. 

\section{Conclusion}
In this work, we identify two loopholes in the WiFi protocol and demonstrate their possible privacy and security threats. In particular, we reveal that today's WiFi radio responds to packets from unauthorized devices outside of the network and it can be easily manipulated to keep awake. These loopholes can be exploited by malicious attackers to jeopardize our daily use of WiFi devices. As examples, we demonstrate how an attacker can take advantage of these loopholes to extract private information such as breathing rate and quickly exhaust the battery of a typical IoT device, leaving the victim's device in a disabled state. 